\documentclass[prb,nofootinbib,twocolumn,superscriptaddress]{revtex4} 

\usepackage{graphicx}
\usepackage{dcolumn}
\usepackage{bm}
\usepackage{threeparttable}
\usepackage{times}
\usepackage{mathptmx}
\usepackage{lscape}
\usepackage{natbib}
\usepackage{amsmath}
\usepackage{amssymb}
\usepackage{braket}
\usepackage{comment}
\usepackage{color}

\def\degree{\kern-.2em\r{}\kern-.3em}

\begin{document}


\title{ Thermodynamics for Nonlinearity under Hidden Structure Information }

\author{Koretaka Yuge}
\affiliation{
Department of Materials Science and Engineering,  Kyoto University, Sakyo, Kyoto 606-8501, Japan\\
}%

\begin{abstract}
{  For substitutional alloys, typically reffered to as classical discrete systems under constant composition, we theoretically examine the role of hidden structure information on evolution of nonliearity (i.e., correspondence between a set of potential energy surface and that of many-body interaction) in canonical ensemble, in terms of the stochastic thermodynamics.  When thermodynamic properties for a given paritial system is controlled by those for e.g., bulk as a hidden structure information, we derive that change in nonlinearity on statistical manifold through any transition is identical to the sum of negative bath entropy change, fluctuation of system entropy change and fluctuation of stochastic mutual information change between the system interested and hidden system: We successfully establish basic formulation of how geometric aspect of nonliearity evolves under feedback from hidden system information, which especially provides deeper insight into the nonlinearity for surface and interaface alloys controlled under bulk thermodynamics. 
}
\end{abstract}


\maketitle

\section{Introduction}
When we consider substitutional multicomponent alloys, it is typicaly reffered to as a 
classical discrete system under \textit{constant} composition. 
In thermodynamic equilibrium, expectation value for structure along chosen coordination $p$ under given coordination $\left\{ q_{1}, \cdots, q_{f} \right\}$ can be given by the following canonical average,
\begin{eqnarray}
\label{eq:can}
\Braket{ q_{p}}_{Z} = Z^{-1} \sum_{i} q_{p}^{\left( i \right)} \exp \left( -\beta U^{\left( i \right)} \right),
\end{eqnarray}
$\beta=\left( k_{\textrm{B}}T \right)^{-1}$ denotes inverse temperature, $Z=\sum_{i}\exp\left(-\beta U_{i}\right)$ partition function,  and summation is taken over all possible microscopic states on configuration space. 
When we employ complete orthonormal basis for e.g., generalized Ising model (GIM),\cite{ce}, potential energy $U$ for any configuration $k$ is exactly expressed by
\begin{eqnarray}
\label{eq:u}
U^{\left( k \right)} = \sum_{j} \Braket{U|q_{j}} q_{j}^{\left( k \right)},
\end{eqnarray}
where $\Braket{\quad|\quad}$ denotes inner product, i.e., trace over possible configurations.  
When we introduce two $f$-dimensional vectors of $\mathbf{q}_{Z}=\left(\Braket{ q_{1}}_{Z},\cdots, \Braket{ q_{f}}_{Z}\right)$ and $\mathbf{U}=\left(U_{1},\cdots,U_{f}\right)$ ($U_{b}=\Braket{U|q_{b}}$), it is clear from 
Eqs.~\eqref{eq:can} and~\eqref{eq:u} that generally, $\mathbf{q}_{Z}$ is a complicted nonlinear function of $\mathbf{U}$, i.e., thermodynamic (here, canonical) average $\phi_{\textrm{th}}$ is a nonlinear map. 

Whereas various approaches have been developed including Metropolis algorism, entropic sampling and Wang-Landau sampling for effective exploration of configuration merged from the nonlinear map to determine equilibrium properties,\cite{mc1,mc2,wl} origin of the nonlinearity in terms of configurational geometry, i.e., geometric information in configuration space without requiring any thermodynamic information such as temperature or energy, has not been well addressed so far.
Our recent theoretical study reveal that nonlinearity can be reasonablly treated through introduced vector field $\mathbf{A}$ of \textit{anharmonicity in s.d.f.} (ASDF) depending only on configurational geometry,\cite{asdf,em2} where $\mathbf{A}$ can be naturally treated as time-evolution of discrete dynamical system. We quantitatively formulate bidirectional stability (BS) character of thermodynamic average between equilibrium structure and potential energy surface in terms of their hypervolume correspondence, by divergence and Jacobian for $\mathbf{A}$,\cite{bd} and we examine the origin of nonliearity based on tropical geometry and information geometry with dually flat Riemannian manifold, which clarifies how spatial constraint to individual s.d.f., entanglement between s.d.f. and nonadditivity of the entanglement dominate the nonlinearity.\cite{trop, ig} 
Following these studies, we recently, further investigate the evolution of nonliearity including its fluctuation in analogy with stochastic thermodynamics, by transforming deterministic dynamical system for evolution of nonlinearity to stochastic Markovian system. 
Then we find that through the transition from one to another state, information gain about the nonlinearity is identical to negative sum of entropy change for thermal bath and fluctuation of stochastic system entropy change from special deterministic transition: Evolution of nonlinearity in thermodynamics on configuration space is bridged to nonlinearity on information geometry. 

Although these studies provide profoundly deeper geometric interpretation of the nonlinearity, its application is confined to a single bulk system, i.e., it cannot be applied to partial (e.g., surface and interface) system under hidden bulk structure information. 
The present study tuckle this problem, and we successfully establish basic relationships between change in nonlinearity through transition, system and bath entropy change, and mutual information between paritial and bulk system. The details are shown below.

\section{Derivation and Concept}
\subsection{Physical Setup}
Let us first briefly explain the concept of local nonlinearity in canonical ensemble defined on individual configuration $\vec{q}$, ASDF, defined as 
\begin{eqnarray}
\label{eq:asdf}
A\left( \vec{q} \right) = \left\{ \phi_{\textrm{th}}\left( \beta \right)\circ \left( -\beta\cdot \Gamma \right)^{-1} \right\}\cdot \vec{q} - \vec{q}.
\end{eqnarray}
Here, $\Gamma$ denotes $f\times f$ covariance matrix for configurational density of states (CDOS) \textit{before} applying many-body interaction to the system, and ASDF corresponds to the vector fieid on configuration space. 
Eq.~\eqref{eq:asdf} indicates that we can retreat nonlinearity of ASDF as a time-evolution of the following discrete dynamical system:
\begin{eqnarray}
\vec{q}_{t+1} = \vec{q}_{t} + A\left( \vec{q}_{t} \right).
\end{eqnarray}

In our previous study, we extend the above deterministic time evolution to stochastic time evolution by introducing the following master equation for probability distribution of $\vec{q}_{n}$:
\begin{eqnarray}
\label{eq:master}
\frac{d}{dt} P\left( \vec{q}_{n} \right) = \sum_{m} \left\{   P\left( \vec{q}_{m} \right) g_{m}\left( \vec{q}_{n} \right)   -  P\left( \vec{q}_{n}\right) g_{n}\left( \vec{q}_{m} \right) \right\}.
\end{eqnarray}
Here, transition probablity from state A to B is given by
\begin{eqnarray}
\label{eq:p}
P \left( \vec{q}_{B} \middle| \vec{q}_{A} \right) = g_{A}\left( \vec{q}_{B} \right) =  \cfrac{g\left( \vec{q}_{B} \right) \exp\left[ -\beta \left( \vec{q}_{B}\cdot \vec{v}_{A} \right) \right]} {\sum_{\vec{q}} g\left( \vec{q} \right) \exp \left[ -\beta \left( \vec{q}\cdot \vec{v}_{A} \right) \right] },
\end{eqnarray}
where $g\left( \vec{q} \right)$ denotes CDOS, and 
\begin{eqnarray}
\vec{v}_{J} := \left( -\beta \cdot \Gamma \right)^{-1} \vec{q}_{J}.
\end{eqnarray}

\begin{figure}[h]
\begin{center}
\includegraphics[width=0.64\linewidth]{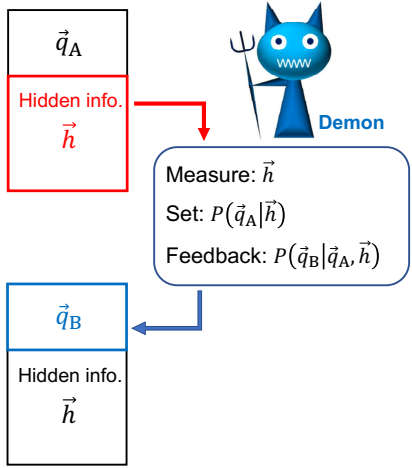}
\caption{Schematic illustration of evolution of nonlinearity, where transition of the system from state A to B is affected by the hidden state $\vec{h}$ with Maxwell's demon feedback. }
\label{fig:demon}
\end{center}
\end{figure}

In the present study, we would like to extend the above stochastic evolution of nonlinearity to including effect of hidden structure information, where the corresponding physical set up is schematically illustrated in Fig.~\ref{fig:demon}.
Here, we first focus on that hidden structure state, $\vec{h}$, is kept fixed for its initial condition, while the system state evolves from A to B: this holds for substitutional alloys where the system corresponds to surface and/or interface and the hidden system to bulk whose thermodynamic state is invariant w.r.t. the state for partial system. With these preparations, we can redefine the transition probablity from A to B under given $\vec{h}$:
\begin{eqnarray}
\label{eq:trans}
P\left( \vec{q}_{B} \middle| \vec{h},\vec{q}_{A} \right) &=& g_{Ah}\left( \vec{q}_{A},\vec{q}_{B},\vec{h} \right) \nonumber \\
&=&\frac{\sum_{\vec{h}} g\left( \vec{Q}_{B} \right) \exp\left[ -\beta\left( \vec{Q}_{B}\cdot \vec{V}_{A} \right) \right]    }{ \sum_{\vec{q},\vec{h}} g\left( \vec{Q} \right)  \exp\left[ -\beta\left( \vec{Q}\cdot \vec{V}_{A} \right) \right]  },
\end{eqnarray}
where
\begin{eqnarray}
\label{eq:trans2}
\vec{Q} :&=& \left( \vec{q},\vec{h} \right) \nonumber \\
\vec{Q}_{K}:&=& \left( \vec{q}_{K}, \vec{h} \right) \nonumber \\
\vec{V}_{K} :&=& \left( -\beta\cdot\Gamma \right)^{-1}\cdot \vec{Q}_{K},
\end{eqnarray}
and we define the denominator as partition function $Z_{A}$, and also introduce corresponding free energy of $F_{A}$.
Note that in Eq.~\eqref{eq:trans2}, $\Gamma$ corresponds to the covariance matrix includig both the system and the hidden system.

In analogy to stochastic thermodynamics, we then define the following thermodynamic functions:
\begin{eqnarray}
\label{eq:s}
\Delta S &=& \ln \frac{g_{Ah}\left( \vec{q}_{A},\vec{q}_{B,}\vec{h} \right)}{g_{Bh}\left( \vec{q}_{A},\vec{q}_{B},\vec{h} \right) } \nonumber \\
\Delta S_{0} &=& \ln \frac{g\left( \vec{q}_{B} \right)}{ g\left( \vec{q}_{A} \right) } \nonumber \\
\Delta S_{G} &=& \ln \frac{g_{GAh}\left( \vec{q}_{A},\vec{q}_{B,}\vec{h} \right)}{g_{GBh}\left( \vec{q}_{A},\vec{q}_{B},\vec{h} \right) } \nonumber \\
\Delta \tilde{S} &=& \Delta S - \Delta S_{G},
\end{eqnarray}
where $S$ denotes bath entropy, $S_{0}$ (initial) system entropy, subscript $G$ represents thermodynamic function for linear system (i.e., CDOS takes Gaussian with the same $\Gamma$ as practical system), and tilde always denotes function measured from that of linear system. 

\subsection{Derivation: Bridge between Geometric and Thermodynamic Nonlinearity}
In our previous study, we successfully bridge change in nonlinearity from state A to B on statistical manifold (i.e., geometric aspect of the nonlinearity) and that of ASDF on configuration space as a thermodynamic function, through taking special operation, called ``vicinity average''. In a similar fashion, we define the followings as the vicinity average under hidden structure information: 
\begin{eqnarray}
\Braket{M}_{AB} &=& \sum_{\vec{q}_{B}} g_{Ah}\left( \vec{q}_{A},\vec{q}_{B},\vec{h} \right)\cdot M  + \sum_{\vec{q}_{A}} g_{Bh}\left( \vec{q}_{A},\vec{q}_{B},\vec{h} \right)\cdot M \nonumber \\
&=& \Braket{M}_{A} + \Braket{M}_{B}.
\end{eqnarray}
We then apply the vicinity average to $\Delta \tilde{S}$, leading to
\begin{widetext}
\begin{eqnarray}
\Braket{\Delta \tilde{S}}_{AB} &=& D_{\textrm{KL}}\left( g_{Ah}: g_{GAh} \right) - D_{\textrm{KL}}\left( g_{Bh} : g_{GBh} \right) - \sum_{\vec{q}_{B}} g_{Ah}\ln\frac{g_{Bh}}{g_{GBh} } + \sum_{\vec{q}_{A}} g_{Bh} \ln\frac{g_{Ah}}{g_{GAh} } \nonumber \\
&=& -\Delta D_{\textrm{KL}}^{A\to B} + \ln\frac{g\left( \vec{q}_{B}, \vec{h} \right)}{ g\left( \vec{q}_{A},\vec{h} \right) } - \ln\frac{g_{G}\left( \vec{q}_{B},\vec{h} \right)}{ g_{G}\left( \vec{q}_{A},\vec{h} \right) } + \beta \Braket{\tilde{F}_{A}}_{B} - \beta\Braket{\tilde{F}_{B}}_{A},
\end{eqnarray}
\end{widetext}
where $D_{\textrm{KL}}$ denotes Kullback-Leibler divergence. 
In order to further clarify the role of hidden system information, we introduce the following functions for practical and linear system, namely 
\begin{eqnarray}
J\left( \vec{q},\vec{h} \right) &=& \frac{g\left( \vec{q},\vec{h} \right)}{ g\left( \vec{q} \right) g\left( \vec{h} \right) } \nonumber \\
J_{G}\left( \vec{q},\vec{h} \right) &=& \frac{g_{G}\left( \vec{q},\vec{h} \right)}{ g_{G}\left( \vec{q} \right) g_{G}\left( \vec{h} \right) },
\end{eqnarray}
which leads that their logarism corresponds to stochastic mutual information, $\ln J \left( \vec{q}_{A}, \vec{h} \right) = i\left( \vec{q}_{A}, \vec{h} \right)$. When we further define the change in stochastic mutual information through transition measured from that for linear system as followings:
\begin{eqnarray}
\Delta \tilde{i} = \left\{ i\left( \vec{q}_{B},\vec{h} \right) - i\left( \vec{q}_{A},\vec{h} \right) \right\} - \left\{ i_{G}\left( \vec{q}_{B},\vec{h} \right) - i_{G}\left( \vec{q}_{A},\vec{h} \right) \right\}, \nonumber \\
\quad
\end{eqnarray}
and when we employ the following characteristics for free energy
\begin{eqnarray}
\Braket{\tilde{F}_{A}}_{AB} &=& \tilde{F}_{A} + \Braket{\tilde{F}}_{B} \nonumber \\
\Braket{\tilde{F}_{B}}_{AB} &=& \tilde{F}_{B} + \Braket{\tilde{F}}_{A},
\end{eqnarray}
we can obtain the relationships between geometric and thermodynamic nonlinearity:
\begin{eqnarray}
\Braket{\Delta\tilde{S} + \beta\tilde{F}}_{AB} = -\Delta D_{\textrm{KL}}^{A\to B} + \Delta \tilde{i} + \beta \Delta \tilde{F}  - \Delta \tilde{S}_{0}.
\end{eqnarray}
From Eq.~\eqref{eq:s} and using the fact that $\Gamma$ is symmetric, we finally obtain the relationship for geometric nonlinearity, entropy change and mutual information under given $\vec{h}$:
\begin{eqnarray}
\label{eq:dkl}
\Delta D_{\textrm{KL}}^{\textrm{A}\to\textrm{B}} =  -\left\{  \Delta\tilde{S}  + \left( 2\Delta\tilde{S}_{0} - \Braket{\Delta\tilde{S}_{0}}_{\textrm{AB}} \right) \right\} + \left( 2\Delta\tilde{i}  - \Braket{\Delta\tilde{i}}_{\textrm{AB}}  \right). \nonumber \\
\quad
\end{eqnarray}
Compared with our previous study, the effect of hidden system information for nonlinearity corresponds to the second term in r.h.s. of 
Eq.~\eqref{eq:dkl}: This fact certainly clarify that when system and hidden system information is separable, there exists no geometric effect of nonlinearity, i.e., the second term takes zero.


\section{Conclusions}
For substitutional alloys under existence of hidden structure information, we clarify that change in nonlinearity on statistical manifold through any transition can always be expressed by the sum of negative bath entropy change, fluctuation of system entropy change and fluctuation of stochastic mutual information change between the system interested and hidden system. The results especially accelerate  deeper understandings of the nonlinearity for surface and interaface alloys controlled by bulk thermodynamics.

\section{Acknowledgement}
This work was supported by Grant-in-Aids for Scientific Research on Innovative Areas on High Entropy Alloys through the grant number JP18H05453 and a Grant-in-Aid for Scientific Research (16K06704) from the MEXT of Japan, Research Grant from Hitachi Metals$\cdot$Materials Science Foundation, and Advanced Low Carbon Technology Research and Development Program of the Japan Science and Technology Agency (JST).

\end{document}